\documentclass[12pt]{article}
\usepackage{amsmath}
\usepackage[dvips]{graphicx}
\usepackage{epsfig}
\usepackage{amsmath}
\usepackage{amssymb}
\usepackage{graphics,epstopdf}
\usepackage{amsfonts}
\usepackage{amsthm}
\usepackage[usenames]{color}

\definecolor{AV}{rgb}{0.65,0.0,0}
\definecolor{GC}{rgb}{0,0.0,0.65}
\definecolor{WS}{rgb}{0,0.65,0}

\usepackage[
      colorlinks=true,
      linkcolor=blue,
      urlcolor=blue,
      filecolor=blue,
      citecolor=red,
      pdfstartview=FitV,
      pdftitle={},
      pdfauthor={},
      pdfsubject={},
      pdfkeywords={},
      pdfpagemode=None,
      bookmarksopen=true
]{hyperref}

\newcommand{\bm}{\begin{multiline}}
\newcommand{\beq}{\begin{equation}}
\newcommand{\eeq}{\end{equation}}
\newcommand{\beqs}{\begin{eqnarray}}
\newcommand{\eeqs}{\end{eqnarray}}

\begin{document}

\thispagestyle{empty}

\hfill{}

\hfill{}

\hfill{}

\vspace{32pt}

\begin{center}

\textbf{\Large On magnetized anisotropic stars}

\vspace{48pt}

\textbf{ Cristian Stelea,}\footnote{E-mail: \texttt{cristian.stelea@outlook.com}}
\textbf{Marina-Aura Dariescu,}\footnote{E-mail: \texttt{marina@uaic.ro}}
\textbf{Ciprian Dariescu, }\footnote{E-mail: \texttt{ciprian.dariescu@uaic.ro}}

\vspace*{0.2cm}

\textit{$^1$ Research Department, Faculty of Physics, ``Alexandru Ioan Cuza" University}\\[0pt]
\textit{11 Bd. Carol I, Iasi, 700506, Romania}\\[.5em]

\textit{$^{2,3}$ Faculty of Physics, ``Alexandru Ioan Cuza" University}\\[0pt]
\textit{11 Bd. Carol I, Iasi, 700506, Romania}\\[.5em]

\end{center}

\vspace{30pt}

\begin{abstract}

We extend a known solution generating technique for isotropic fluids in order to construct more general models of anisotropic stars with poloidal magnetic fields. In particular, we discuss the magnetized versions of some well-known exact solutions describing anisotropic stars and dark energy stars and we describe some of their properties.
\end{abstract}

\vspace{32pt}

\setcounter{footnote}{0}

\newpage

\section{Introduction}

There are strong theoretical reasons to believe that in realistic relativistically covariant stellar models in the high density regimes the pressures inside the star are likely to be anisotropic \cite{Ruderman:1972aj}. This usually means that the radial pressure component, $p_r$ is not equal to the components in the transverse directions, $p_t$. Such anisotropies in the fluid distributions can arise from various reasons: it can be due to a mixture of two fluid components \cite{Letelier:1980mxb}, the existence of a superfluid phase, the presence of a magnetic field, etc. (for a review see \cite{Herrera:1997plx} and references there). Other non-trivial examples of anisotropic fluid distributions are provided by the bosonic stars (see for instance \cite{Liebling:2012fv} and the references within), by traversable wormholes \cite{Bronnikov:2017kvq}, or the so-called gravastars \cite{Cattoen:2005he}, which are systems where anisotropic pressures occur naturally.

The pressure anisotropy can have significant effects on the structure and the properties of the stellar models. For instance, for an anisotropic star with mass $M$ and radius $R$, the quantity $\frac{2M}{R}$ can approach unity \cite{Karmakar:2007fn}, while it is constrained in the isotropic models by the Buchdahl bound $\frac{2M}{R}<\frac{8}{9}$ \cite{Buchdahl}. Also, the surface redshift of the star can be arbitrarily large (although it is still bounded when one takes into account the energy conditions inside the star \cite{Ivanov:2002xf}). 

Such anisotropic fluid models of neutron stars could be used to model the so-called magnetars \cite{DT}, which denote a class of neutron stars whose emissions are powered by the decay of their huge magnetic field. For a magnetar the magnetic field strength can reach values as high as $10^{11}T$, while being even more intense inside the star. There are over $30$ magnetars cataloged by now \cite{Olausen:2013bpa} (for recent reviews of their properties see \cite{Esposito:2018gvp}, also \cite{Woods:2004kb}). This class of objects includes the soft gamma repeaters (SGRs) and the the anomalous X-ray pulsars (AXPs). To model such objects in fully relativistic context is quite a formidable task. As expected, the strong magnetic field of the magnetars has significant effects on the structure of a neutron star. Usually, in modeling compact objects in General Relativity (GR) one assumes small deviations from the spherical symmetry and a perfect fluid distribution of the source (see for instance \cite{Colaiuda:2007br}). However, in presence of strong magnetic fields one should consider an axially symmetric treatment of the source \cite{Negreiros:2018cjk}. 

One simple non-perturbative model of a magnetar, which is a solution of the full Einstein-Maxwell-hydrodynamic equations has been proposed recently by Yazadjiev in \cite{Yazadjiev:2011ks}. Yazadjiev's model describes a static magnetized neutron star, with a poloidal magnetic field. Its fluid distribution becomes anisotropic due to the presence of the magnetic field. This model was obtained using a solution generating technique by adding a magnetic field to a general static metric, solution of the Einstein-perfect fluid model. This solution is very important since it provides in a fully relativistic context a simple analytical model of the magnetars.

In our work we shall generalize Yazadjiev's method by allowing the seed solution to describe an anisotropic fluid, instead of a perfect fluid. In other words, in our solution the pressure anisotropy exists even before `turning on' the magnetic field. The presence of the magnetic field has the further effect to make the anisotropy complete, in the sense that the pressures in the transverse directions, $p_{\theta}$ and $p_{\phi}$ become non-equal. Thus, in our final solution all the pressures (along the radial  and transverse directions) become different. While the condition $p_{\theta}=p_{\phi}$ is necessarily obeyed for a system with spherical symmetry, in our case the final geometry has only axial symmetry and the two transverse pressures are then allowed to differ. 

The structure of our paper is as follows: in the next section we present in simple form the extension of the Yazadjiev's procedure to add a magnetic field to a static configuration describing an anisotropic fluid. In section $3$ we consider some exact solutions to illustrate the procedure and study some of their properties. Finally, the last section is dedicated to conclusions.  

\section{The magnetized model}

As the starting point of our solution-generating technique, we shall consider a static spherically symmetric anisotropic distribution of matter for which the metric can be written in general as:
\beqs
\label{seed}
ds^2&=&-g_{tt}(r)dt^2+g_{rr}(r)dr^2+r^2(d\theta^2+\sin^2\theta d\varphi^2).\label{seed}
\eeqs
It describes a solution of the Einstein-anisotropic fluid equations:\footnote{Note that we work in the natural units for which $G=c=1$.}
\beqs 
G_{\mu\nu}&=&8\pi T_{\mu\nu}^0,
\eeqs 
where the stress-energy $T_{\mu\nu}^0$ of the anisotropic distribution of matter has the form:
\beqs
T_{\mu\nu}^0&=&(\rho^0+p_t^0)u_{\mu}^0u_{\nu}^0+p_t^0g_{\mu\nu}^0+(p_r^0-p_t^0)\chi_{\mu}^0\chi_{\nu}^0.
\eeqs
Here $\rho^0$ is the fluid density, $p_r^0$ is the radial component of the pressure, while $p_t^0$ represents the transverse components of the pressure. Moreover, $u_{\mu}^0$ is the $4$-velocity of the fluid while $(\chi^0)^{\mu}=\sqrt{g_{rr}^{-1}}\delta^{\mu}_r$ is the unit spacelike vector in the radial direction.

Following the same reasoning as in the original Yazadjiev's method \cite{Yazadjiev:2011ks}, starting from the spherically symmetric seed (\ref{seed}) the final magnetized solution becomes:
\beqs
\label{final}
ds^2&=&\Lambda^2\bigg[-g_{tt}(r)dt^2+g_{rr}(r)dr^2+r^2d\theta^2\bigg]+\frac{r^2\sin^2\theta}{\Lambda^2}d\varphi^2,~~~ \Lambda=1+\frac{B_0^2}{4}r^2\sin^2\theta,\nonumber\\
A_{\varphi}&=&\frac{B_0}{2}\frac{r^2\sin^2\theta}{\Lambda},~~~~~\rho=\frac{\rho^0}{\Lambda^2},~~~ p_{\theta}=\frac{p_t^0}{\Lambda^2},~~~p_r=\frac{p_r^0}{\Lambda^2},~~~j_{\varphi}=(\rho-p_r)\frac{B_0r^2\sin^2\theta}{\Lambda^2},\label{finalmag}
\eeqs
which is a solution of the Einstein-Maxwell-anisotropic fluid equations:
\beqs
G_{\mu\nu}&=&8\pi T^{EM}_{\mu\nu}+8\pi T^{af}_{\mu\nu},~~~~F^{\mu\nu}_{;\nu}=4\pi J^{\mu}.\label{EMAI}
\eeqs
Here $F=dA$ is the Maxwell field strength with the magnetic potential $A=A_{\varphi}d\varphi$, $J_{\mu}=(0, 0, 0, j_{\varphi})$ is the $4$-current that sources the electromagnetic field, while the stress-energy tensor of the electromagnetic field is defined as usual by:
\beqs
T^{EM}_{\mu\nu}&=&\frac{1}{4\pi}\left(F_{\mu\alpha}F^{~\alpha}_{\nu}-\frac{1}{4}F^2g_{\mu\nu}\right).
\eeqs
The final anisotropic fluid configuration is described by means of the stress-energy tensor:
\beqs
T^{af}_{\mu\nu}&=&(\rho+p_t)u_{\mu}u_{\nu}+p_tg_{\mu\nu}+(p_r-p_t)\chi_{\mu}\chi_{\nu}+\sigma\zeta_{\mu}\zeta_{\nu},
\eeqs
which manifestly includes the anisotropic contribution due to the presence of the magnetic field:
\beqs
\sigma&=&-(\rho-p_r)\frac{B_0^2r^2\sin^2\theta}{2\Lambda},
\eeqs
where $\zeta_{\mu}$ is the unit vector along the $\frac{\partial}{\partial\varphi}$ direction. 

Note that the anisotropy induced by the magnetic field is directed solely along the $\zeta_{\mu}$ direction and that it breaks the condition $p_{\varphi}=p_{\theta}=p_t$. This condition was required by the assumed spherical symmetry in the seed solution. However, in presence of the magnetic field the geometry is no longer spherically symmetric and as such one can have the two transverse pressures non-equal, $p_{\varphi}=p_{\theta}+\sigma\leq p_{\theta}$.

\section{Some anisotropic examples}

Starting with the pioneering work of Bowers and Liang \cite{Bowers}, there has been a growing interest in developing more realistic stellar models by using anisotropic fluid distributions. A relatively recent review of the properties of compact objects with anisotropic fluids and the possible causes of for the appearance of local pressure anisotropies is \cite{Herrera:1997plx}. For more recent papers on anisotropic compact objects see for instance \cite{Harko:2002db} - \cite{Herrera:2007kz} and also \cite{Ivanov:2017kyr} and references within. 

A physical anisotropic solution for a stellar model has to fulfill the following requirements (see for instance \cite{Mak:2001eb}
, \cite{Ivanov:2017kyr}):
\begin{itemize}
\item $\rho$, $p_r$ and $p_t$ should all be positive inside the object and finite at the center of the star. Moreover, the anisotropy should be zero at the origin, $p_t(0)=p_r(0)$.
\item The gradients of  $\rho$, $p_r$ and $p_t$ should be negative. These means that the density and pressures should reach a maximum at the center at the star and decrease radially outwards. The transverse pressures should be bigger than the radial pressure, except at the origin, $p_t>p_r$.
\item The sound speeds should be bounded by that of light:
\beqs
0<\frac{\partial p_r}{\partial\rho}<1,~~~~~0<\frac{\partial p_t}{\partial\rho}<1,
\eeqs
\item For stability the adiabatic index $\Gamma=\frac{\rho+p_r}{p_r}\frac{dp_r}{\rho}$ should be larger than $\frac{4}{3}$. 
\item The solution should satisfy the energy conditions inside the star. For instance, the dominant energy condition is $\rho\geq p_r$ and $\rho\geq p_t$.
\item Stability against cracking, which for spherical symmetry requires that
\beqs
-1&\leq&\frac{dp_t}{d\rho}-\frac{dp_r}{d\rho}\leq 0
\eeqs
\end{itemize}
In the followings we shall consider bounded configurations of an anisotropic fluid for which the radial pressure $p_r$ will vanish at a finite radius $R$, which defines the surface of the star. Moreover, at the surface of the star $r=R$ the interior anisotropic solution should match continously the exterior Schwarzschild solution:
\beqs
ds^2&=&-\left(1-\frac{2M}{R}\right)dt^2+\frac{dr^2}{1-\frac{2M}{R}}+r^2(d\theta^2+\sin^2\theta d\varphi^2).\label{sch}
\eeqs
Here $M$ is the mass of the star. One should note that the density $\rho$ and the transverse pressure $p_t$ are not required to vanish at the surface of the star. 

Since most of these models assume spherical symmetry they can be used as seeds in the method described in the previous section. Here we shall consider, as examples, a few representative anisotropic solutions.

\subsection{The magnetized Bowers-Liang solution}

The solution found by Bowers and Liang \cite{Bowers} corresponds to a anisotropic fluid with an homogeneous density distribution $\rho=\rho_0=constant$.  In their work they considered a spherically symmetric relativistic matter distribution and studied the behavior of such systems by incorporating the pressure anisotropy effects in the equation of the hydrostatic equilibrium. Their solution is given by (\ref{seed}) where:
\beqs
g_{tt}(r)&=&\bigg[\frac{3\left(1-\frac{2M}{R}\right)^{\frac{h}{2}}-\left(1-\frac{2m(r)}{r}\right)^{\frac{h}{2}}}{2}\bigg]^{\frac{2}{h}},~~~~g_{rr}(r)=\frac{1}{1-\frac{2m(r)}{r}},~~~\rho^0=\frac{3M}{4\pi R^3},\\
p_r^0&=&\rho^0\frac{\left(1-\frac{2m(r)}{r}\right)^{\frac{h}{2}}-\left(1-\frac{2M}{R}\right)^{\frac{h}{2}}}{3\left(1-\frac{2M}{R}\right)^{\frac{h}{2}}-\left(1-\frac{2m(r)}{r}\right)^{\frac{h}{2}}},~~~\Delta^0=p_t^0-p_r^0=\frac{4\pi}{3}Cr^2\frac{(\rho^0+p_r^0)(\rho^0+3p_r^0)}{1-\frac{2m(r)}{r}},\nonumber
\eeqs
where $h=1-2C$, $m(r)=\frac{4\pi}{3}r^3\rho^0$ and $C$ is the anisotropy parameter.

Then, using the results from section $2$ the magnetized Bowers-Liang solution will simply be given by (\ref{finalmag}). The final geometry is obviously axially symmetric, while the anisotropic fluid source has the pressures different in radial and transverse directions. Note also that for $C=0$ one obtains the magnetized interior Schwarzschild solution discussed in \cite{Yazadjiev:2011ks}. Since in origin $\Lambda=1$, then the magnetized solution has the same values for the density and pressures as the original seed at $r=0$. In particular, the radial pressure in origin becomes:
\beqs
p_r(0)&=&\rho^0\frac{1-\left(1-\frac{2M}{R}\right)^{\frac{h}{2}}}{3\left(1-\frac{2M}{R}\right)^{\frac{h}{2}}-1},
\eeqs
and the critical value of the quantity $\frac{2M}{R}$ for which the central pressure becomes infinite is the same as in the original Bowers-Liang solution:
\beqs
\frac{2M}{R}|_{cr}&=&1-\left(\frac{1}{3}\right)^{\frac{2}{h}}.
\eeqs

If one takes $h=0$ in the magnetized Bowers-Liang solution one obtains the magnetized version of the so-called Florides solution \cite{florides}. It corresponds to an anisotropic object with zero radial pressure $p_r=0$, which is sustained only by tangential stresses. In this case:
\beqs
g_{tt}(r)&=&\frac{\left(1-\frac{2M}{R}\right)^{\frac{3}{2}}}{\left(1-\frac{2m(r)}{r}\right)^{\frac{1}{2}}},~~~g_{rr}(r)=\frac{1}{1-\frac{2m(r)}{r}},~~~\rho=\frac{1}{\Lambda^2}\frac{3M}{4\pi R^3},~~~m(r)=M\frac{r^3}{R^3}\nonumber\\
p_{\theta}&=&\frac{2\pi\rho^2r^2}{3\left(1-\frac{2m(r)}{r}\right)},~~~p_{\varphi}=p_{\theta}+\sigma,~~~p_r=0.
\eeqs

Since the effects of the magnetic field are manifest not only in the interior of the star and also outside, the proper exterior geometry to consider outside the star's surface at $r=R$ is the Schwarzschild-Melvin solution. It corresponds to $g_{tt}=g_{rr}^{-1}=1-\frac{2M}{R}$ in (\ref{finalmag}), while the density all the fluid pressures are zero. Since the interior Bowers-Liang matches continuously the exterior Schwarzschild vacuum solution on the surface $r=R$, it should be obvious that the magnetized Bowers-Liang solution will match continuously the exterior Schwarzschild-Melvin solution on the star's surface. 

Finally, one should note that some of the regularity conditions described in this section have been derived in a spherically symmetric geometry and some of them might not be appropriate in discussing anisotropic fluid distributions with general axial symmetry. We leave this point outside of this work and we will address it elsewhere.

\subsection{Magnetized dark energy stars}

The second example that we shall discuss is provided by the so-called dark energy stars. These objects were originally defined by Mazur and Mottola \cite{Mazur:2001fv}, \cite{Mazur:2004ku}, \cite{Mazur:2004fk} as `gravastars' (gravitational vacuum stars) and they were introduced as alternatives to black holes. They usually contain a de Sitter core surrounded by a finite layer of stiff matter with the equation of state $p=\rho$, which is also matched with an exterior Schwarzschild vacuum geometry. By construction, this object has no singularity at the origin and no horizon, while its radius can be made very close to that of the Schwarzschild radius. Moreover, for some physically reasonable equations of state for the matter in the transition layers they can be dynamically stable against spherically symmetric perturbations of the matter or gravity fields \cite{Visser:2003ge}. Since their exterior geometry coincides with that of a Schwarzschild black hole, it might be very difficult to observationally distinguish a gravastar from a Schwarzschild black hole. However, one should note that gravitational waves might be able to distinguish gravastars from black holes \cite{Pani:2009ss}.

In this section we shall use a generalized model of a gravastar found by Lobo and described in \cite{Lobo:2005uf}. In this case the interior de Sitter core is replaced by an anisotropic dark energy fluid having the equation of state of the form $p_r=\omega \rho$, with $\omega<-\frac{1}{3}$. For simplicity we shall consider here the case of constant energy density $\rho^0(r)=\rho^0=constant$.

The seed solution that we are considering is then the following:
\beqs
ds^2&=&-(1-2Ar^2)^{-\frac{1+3\omega}{2}}dt^2+\frac{dr^2}{1-2Ar^2}+r^2(d\theta^2+\sin^2\theta d\varphi^2),\nonumber\\
\rho^0&=&\frac{3A}{4\pi},~~~p_r^0=\omega\rho^0, ~~~p_t^0=p_r^0\big[1+\frac{(1+\omega)(1+3\omega)Ar^2}{2\omega(1-2Ar^2)}\big].
\eeqs
 One is now ready construct the magnetized dark energy star:
\beqs
ds^2&=&\Lambda^2\bigg[-(1-2Ar^2)^{-\frac{1+3\omega}{2}}dt^2+\frac{dr^2}{1-2Ar^2}+r^2d\theta^2\bigg]+\frac{r^2\sin^2\theta}{\Lambda^2}d\varphi^2,~~~ \Lambda=1+\frac{B_0^2}{4}r^2\sin^2\theta,\nonumber\\
A_{\varphi}&=&\frac{B_0}{2}\frac{r^2\sin^2\theta}{\Lambda},~~~\rho=\frac{\rho^0}{\Lambda^2},~~~p_{\theta}=\frac{p_t^0}{\Lambda^2},~~~p_r=\omega\rho,~~~p_{\varphi}=p_{\theta}-(\rho-p_r)\frac{B_0^2r^2\sin^2\theta}{2\Lambda}.
\eeqs
This is a solution of the Einstein-Maxwell-anisotropic fluid equations (\ref{EMAI}) where
\beqs
j_{\varphi}=(\rho-p_r)\frac{B_0r^2\sin^2\theta}{\Lambda^2}
\eeqs

For a static star one should match this geometry with that of the Schwarzschild-Melvin solution on the junction surface $r=R$. As it is well known, there is a set of regularity conditions to be imposed, namely the Israel-Lanczos junction conditions \cite{Israel:1966rt}, \cite{lanczos}. In our case the surface $r=R$ has the induced metric:
\beqs
ds^2&=&h_{ab}dx^adx^b=\Lambda_0^2\bigg[-\left(1-\frac{2M}{R}\right)dt^2+R^2d\theta^2\bigg]+\frac{R^2\sin^2\theta}{\Lambda_0^2}d\varphi^2, ~~~\Lambda_0=1+\frac{B_0^2}{4}R^2\sin^2\theta.\nonumber
\eeqs
The continuity of the metric at this junction leads to the following condition, connecting the two parameters $M$ and $A$:
\beqs
1-\frac{2M}{R}&=&(1-2AR^2)^{-\frac{1+3\omega}{2}}.
\eeqs
Finally, the surface stress-energy tensor $S^a_b$ at the junction surface $r=R$ is given by the Lanczos equations:
\beqs
S^a_b&=&-\frac{1}{8\pi}(k^a_b-\delta^a_b k^c_c),
\eeqs
where $k_{ab}=K_{ab}^{+}-K_{ab}^{-}$ is the discontinuity in the second fundamental form of the junction surface. Here the extrinsic curvature is defined as $K_{ab}=h_a^ch_b^d \chi_{d;c}$, where $\chi$ is the radial unit vector, normal to the surface $r=R$.

For the exterior Schwarzschild-Melvin geometry one finds:
\beqs
K^t_t&=&\frac{\frac{M}{R^2}}{\sqrt{1-\frac{2M}{R}}}\frac{1}{\Lambda_0}+\sqrt{1-\frac{2M}{R}}\frac{1}{\Lambda_0^2}\frac{\partial \Lambda}{\partial r}|_{r=R},\nonumber\\
K^{\theta}_{\theta}&=&\frac{\sqrt{1-\frac{2M}{R}}}{\Lambda_0^2}\left(\frac{\Lambda_0}{R}+\frac{\partial\Lambda}{\partial r}|_{r=R}\right),\nonumber\\
K^{\varphi}_{\varphi}&=&\frac{\sqrt{1-\frac{2M}{R}}}{\Lambda_0^2}\left(\frac{\Lambda_0}{R}-\frac{\partial\Lambda}{\partial r}|_{r=R}\right).
\eeqs
A similar computation for the interior magnetized dark energy star leads to:
\beqs
K^t_t&=&\frac{A(1+3\omega)R}{\sqrt{1-2AR^2}}\frac{1}{\Lambda_0}+\sqrt{1-2AR^2}\frac{1}{\Lambda_0^2}\frac{\partial \Lambda}{\partial r}|_{r=R},\nonumber\\
K^{\theta}_{\theta}&=&\frac{\sqrt{1-2AR^2}}{\Lambda_0^2}\left(\frac{\Lambda_0}{R}+\frac{\partial\Lambda}{\partial r}|_{r=R}\right),\nonumber\\
K^{\varphi}_{\varphi}&=&\frac{\sqrt{1-2AR^2}}{\Lambda_0^2}\left(\frac{\Lambda_0}{R}-\frac{\partial\Lambda}{\partial r}|_{r=R}\right).
\eeqs

Then the surface stresses computed from the Lanczos equations are:
\beqs
\sigma&=&\frac{1}{4\pi R\Lambda_0}\bigg[\sqrt{1-\frac{2M}{R}}-\sqrt{1-2AR^2}\bigg],\nonumber\\
{\cal P}_{\theta}&=&\frac{1}{8\pi R\Lambda_0}\bigg[\frac{1-\frac{M}{R}}{\sqrt{1-\frac{2M}{R}}}-\frac{1-AR^2(1-3\omega)}{\sqrt{1-2AR^2}}\bigg],\nonumber\\
{\cal P}_{\varphi}&=&{\cal P}_{\theta}+\frac{R}{\Lambda_0}\frac{\partial\Lambda}{\partial r}|_{r=R}\sigma,
\eeqs
where $\frac{\partial\Lambda}{\partial r}|_{r=R}=\frac{B_0R\sin^2\theta}{2}$. Here $\sigma$ is the surface energy density, while ${\cal P}_{\theta}$ and ${\cal P}_{\varphi}$ are the tangential surface pressures. Note that the matter distribution on the thin-shell becomes anisotropic due to the presence of the magnetic field.

A more general discussion of the gravitational stability of the dark energy magnetar could be carried out by allowing the junction surface shell to move radially, however, this analysis is outside the purpose of the present work.

\section{Conclusions}

In the last decades there has been a tremendous interest in finding new exact solutions of Einstein's field equations, in particular solutions that describe relativistic compact objects. With the discovery of many new interesting astrophysical objects, such as pulsars, neutron stars, magnetars, etc. one had to increase the level of sophistication in models describing such objects to take into account their observed properties. Although in order to realistically analyze such complicated stellar structures one has to rely on perturbative methods or on complex numerical computations in GR, sometimes, some simple toy models can capture important features of these fascinating objects. For example, one simple nonperturbative model of magnetars has been recently introduced by Yazadjiev in \cite{Yazadjiev:2011ks}. In this work we have considered an extension of Yazadjiev's method to magnetize an anisotropic fluid distribution in full generalyl relativistic context. Our initial motivation was to search for more general models of magnetars with poloidal magnetic fields and in this regard the extension of Yazadjiev's method to anisotropic fluid distributions is quite natural. Indeed in a realistic star model the anisotropy can have many sources and, therefore, in general one should take them into account even in absence of a ultra-strong magnetic field. One of the results of our paper was that the anisotropy induced by the presence of the magnetic field has the effective effect of making all the pressures non-equal along the different directions inside a relativistic magnetized star. This is allowed once we evade the constraints of spherical symmetry, which forces the two transverse pressures $p_{\theta}$ and $p_{\varphi}$ to be equal. As examples of this method we presented two new magnetized versions of some well-known anisotropic solutions: the Bowers-Liang solution and the dark energy stars. The geometry outside the magnetized stars can be well described by the Schwarzschild-Melvin solution and on the surface of the star one has to match continuously the interior geometry with the outside one. In the case of the dark energy star we explicitly computed the surface stresses on the thin-shell separating the interior geometry to that of the exterior Schwarzschild-Melvin solution. A more complete discussion of the stability of the dark energy star should involve dynamic thin-shells, which could move radially, however we leave such a discussion for further work.

As avenues for further work, one should be able to extend this magnetizing method to more general axially-symmetric geometries as required for realistic magnetars in \cite{Negreiros:2018cjk}. These geometries will likely require more general anisotropic fluid interior solutions with axial symmetry as in \cite{Hernandez-Pastora:2016ctg}, \cite{Herrera:2013hm}. Another interesting extension of the present work would involve a study of the star's anisotropy on the propagation of various fields in this background, on the lines of the study presented in \cite{Dariescu:2017ima}. Work on these matters is in progress and it will be presented elsewhere.

\vspace{10pt}

{\Large Acknowledgements}

This work was financially supported by a grant of Ministery of Research and Innovation, CNCS - UEFISCDI, project number PN-III-P4-ID-PCE-2016-0131.

\end{document}